\documentclass[epj]{svjour}

\usepackage{graphicx}
\usepackage{amsmath}
\usepackage{amssymb}
\newcommand{\putfig}[4]{\begin{figure}\begin{center}\includegraphics[height=#3]{#2}\caption{#4\label{#1}}\end{center}\end{figure}}
\newcommand{\spanfig}[4]{\begin{figure*}\begin{center}\includegraphics[height=#3]{#2}\caption{#4\label{#1}}\end{center}\end{figure*}}

\newcommand{\scS}{\mathcal{S}}
\newcommand{\scH}{\mathcal{H}}
\newcommand{\scD}{\mathcal{D}}

\newcommand{\vq}{\mathbf{q}}
\newcommand{\vk}{\mathbf{k}}
\newcommand{\vK}{\mathbf{K}}
\newcommand{\vV}{\mathbf{V}}
\newcommand{\vecr}{\mathbf{r}}

\newcommand{\al}{\alpha}
\newcommand{\alp}{\alpha^{\prime}}
\newcommand{\apr}{b}
\newcommand{\ellp}{\ell^{\prime}}
\newcommand{\qs}{\vq_{S}}

\newcommand{\psum}[2]{\sum_{#1}^{{\,\,\,\,\,\, #2}_{\,\,\scriptstyle{\prime}}}}
\newcommand{\eff}{\textrm{eff}}

\newcommand{\lan}{\left\langle}
\newcommand{\ran}{\right\rangle}

\newcommand{\anav}[1]{\left\langle#1\right\rangle_{\!0}}
\newcommand{\lganav}[1]{\left\langle#1\right\rangle_{\!\!0}}
\newcommand{\quav}[1]{\overline{#1}^{0}}

\newcommand{\effanav}[2]{\left\langle#1\right\rangle_{#2}^{\!\textrm{eff}}}

\begin{document}

\title{Microphase separation in cross-linked polymer blends}
\subtitle{Efficient replica RPA post-processing of simulation data for homopolymer networks}
\author{A.~V.~Klopper\inst{1} \and Carsten~Svaneborg\inst{2} \and Ralf~Everaers\inst{3}}

\institute{Max Planck Institute for the Physics of Complex Systems, N{\"o}thnitzer Stra{\ss}e 38, D-01187 Dresden, Germany \and Department of Chemistry and Interdisciplinary Nanoscience Center (iNano), University of Aarhus, Langelandsgade 140, DK-8000 {\AA}rhus, Denmark \and Universit{\'e} de Lyon, Laboratoire de Physique, {\'E}cole Normale Sup{\'e}rieure de Lyon, CNRS UMR 5672, 46 all{\'e}e d'Italie, 69364 Lyon Cedex 07, France}

\date{\today}

\abstract{
We investigate the behaviour of randomly cross-linked (co)polymer blends using a combination of replica theory and large-scale molecular dynamics simulations. In particular, we derive the analogue of the random phase approximation for systems with quenched disorder and show how the required correlation functions can be calculated efficiently. By post-processing simulation data for homopolymer networks we are able to describe neutron scattering measurements in heterogeneous systems without resorting to microscopic detail and otherwise unphysical assumptions. We obtain structure function data which illustrate the expected microphase separation and contain system-specific information relating to the intrinsic length scales of our networks.}
\PACS{
      {61.25.Hq}{polymer solutions}   \and
      {61.43.Ðj}{disordered solids}   \and
      {64.75.+g}{phase separation}   \and
      {05.40.Ða}{fluctuation phenomena in statistical physics}
     }

\maketitle

\section{Introduction}
\label{Intro}
Highly concentrated liquids comprising long polymeric chains can undergo a process of cross-linking, resulting in the formation of a disordered solid with a frozen memory of its preparation conditions. The cross-linking creates quenched connective and topological disorder which plays a role familiar from glassy systems in breaking the translational symmetry of the initial liquid state. In the case of polymer blends, it prohibits macroscopic phase separation so that demixing can only occur on a microscale \cite{pg-dg79a}. The resulting microstructure is intimately connected with the macroscopic elastic properties of the solid and proves essential to the correct interpretation of neutron scattering experiments on dense cross-linked heterogeneous melts \cite{bb88}. For example, partial deuteration (or `labeling') is often employed in scattering investigations of polymer blends \cite{eb97,mw07}. This process can result in weak but measurable effective interactions between labeled chain sections, inducing a microstructure which influences the system's dynamics and response to strain.

Simulations offer a powerful means by which to compare with the results of such experiments, claiming an additional advantage in complete characterisation and control of the microscopic state \cite{kg95}. In particular, they are capable of isolating (and thus negating) effects due to inhomogeneities in the system prior to cross-linking. In this way, one can use simulations to investigate systems for which heterogeneity does not play a role in the distribution of cross-links.

An approach aimed at circumventing the huge computational expense of simulating blends subject to quenched disorder involves devising a theoretical framework within which heterogeneity may be incorporated after cross-linking, without relying on the exhaustive computation required in brute-force simulations (see e.g. Ref.~\cite{lsb00}). In this respect, a random phase approximation (RPA) for the monomeric density fluctuations is a convenient tool \cite{pg-dg79b}. However, one must take care to modify the standard RPA expression often used to analyse scattering experiments, before applying it to the system with quenched disorder \cite{wuprs96,wprsem01}. The modifications arise from the fact that the density variables fluctuate about some non-zero mean determined by the disorder, which effectively correlates fluctuations associated with different wave vectors, necessitating an average over quenched disorder.

We construct an appropriate formalism, with the view to examine microphase separation due to interactions between components in the melt, introduced after formation of the network. We utilise extensive foundations laid previously for replica formalism in microscopic models \cite{bv92a,bv92b} and field theories \cite{wzg05,wgz06}. Whilst our theory follows in the spirit of these studies, it does so without encoding details about the microscopic structure of the chains. This omission renders it applicable to our simulation data which in turn provides insight into the effects of system-specific parameters on the outcome of (simulated) scattering experiments. Herein lies the strength of this work; we sample correlation functions for density fluctuations in homogeneous networks representing a broad range of network types and cross-linking densities \cite{sge05b,sge04}. From these we can infer the behaviour of heterogeneous cross-linked melts using our RPA formalism. In addition, we have developed an inexpensive means of sampling the correlation functions in randomly labeled homogeneous systems. We can exploit this combination in order to address previously unresolved questions, such as the exact behaviour of the structure factor near microphase separation \cite{pg-dg79a,bb88,rbm95}.

In the next section, we define the system simulated via the molecular dynamics of cross-linked polymer systems and describe in detail the method by which we take care of the quenched disorder.  This section is followed by an application of spin-glass replica formalism \cite{by86} to our data in order to describe neutron scattering measurements in heterogeneous systems. We conclude with a discussion of microphase behaviour and its implications.

\section{Simulated scattering from homogeneous networks}
\label{Sec:1}
We study data from extensive molecular dynamics simulations of randomly cross-linked bead-spring polymer networks \cite{sge05b}. For completeness, we briefly characterise the system in an Appendix. Partial deuteration of the networks is implemented by randomly distributing labels of variable length, $N_{\ell}$, on precursor chains. Of the total number of all available labels, $N_{\al}$, we label a fraction, $\phi$, such that the system contains $\phi N_{\al}$ labels in all. In this way, we can assume there are no correlations between labeled and unlabeled segments. This plays an essential role in simplifying our analysis and renders the calculation of correlation functions far more straightforward than the analogous process for an end-linked triblock copolymer melt, for example.

We obtain correlation functions corresponding to measurements from scattering experiments in terms of label density, $\rho(\vecr)$, which is defined by the position of monomer $\ell$ on label $\alpha$ such that, in Fourier space,
\begin{equation}\label{DefineFourierDensity}
\rho_{\vq}
=
\sum_{\al}^{\phi N_{\al}}\sum_{\ell}^{N_{\ell}}e^{i\vq\cdot\vecr_{\ell}^{\al}}.
\end{equation}
The scattering observed in the precursor melt is then described by the structure function,
\begin{equation}\label{DefineBareStructureFunction}
\scS_{0}(\vq,\vk)
=
\frac{1}{N_{\al}N_{\ell}}\anav{\rho_{\vq}\rho_{\vk}^{\star}},
\end{equation}
where $\anav{\dots}$ represents a thermal average over annealed variables in the context of the molecular dynamics simulations and the subscript `$0$' reminds us that the system is homogeneous. 

The quenched disorder creates correlations between density fluctuations associated with different wave vectors. The translational invariance of the simulated precursor state is lost upon cross-linking and the density variables fluctuate about some non-zero mean determined by the disorder,
\begin{center}
\begin{tabular}{rcc}
$\anav{\rho_{\vq}}\ne0$
& for 
& $\label{NVC1}\vq\ne\mathbf{0}$,
\\
$\anav{\rho_{\vq}\rho_{\vk}^{\star}}\ne0$
& for
& $\vq\ne\vk$.
\end{tabular}
\end{center}
An average over disorder restores the invariance expected in macroscopic systems, and resulting correlation functions may be expressed as,
\begin{subequations}\label{DisorderAveragedCorrelations}
\begin{align}
\quav{\anav{\rho_{\vq}}}
&=
0,\label{DAC1}\\
\quav{\anav{\rho_{\vq}\rho_{\vk}^{\star}}}
&=
N_{\al}N_{\ell}\scS_{0}(\vq)\delta_{\vq\vk},\label{DAC2}\\
\quav{\anav{\rho_{\vq}}\anav{\rho_{\vk}^{\star}}}
&=
N_{\al}N_{\ell}\Gamma_{0}(\vq)\delta_{\vq\vk},\label{DAC3}
\end{align}
\end{subequations}
for $\vq\ne\mathbf{0}$, where the overline denotes a quenched disorder average. These equations describe respectively the vanishing mean of density fluctuations, the structure factor for the homogeneous system and the overlap function familiar from spin glass physics \cite{ea75}. The latter term in particular arises from the broken translational symmetry in the quenched system.

A straightforward way to calculate these averages would be to (randomly) assign fixed labels to chains and to average over a (suitably large) number of independent realisations of the vulcanisation process. In this case, we would need sum over the fraction of labeled monomers for all realisations,
\begin{subequations}\label{SumDisorderRealisations}
\begin{align}
\scS_{0}(\vq)\delta_{\vq\vk}
&=
\frac{X^{-1}}{N_{\al}N_{\ell}}\sum_{y}^{X}\left(\sum_{\al\alp}^{\phi N_{\al}}\lganav{\sum_{\ell\ellp}^{N_{\ell}} e^{i(\vq\cdot\vecr_{\ell}^{\al}-\vk\cdot\vecr_{\ellp}^{\alp})}}\right)_{\!\!y},\label{SDR1}\\
\Gamma_{0}(\vq)\delta_{\vq\vk}
&=
\frac{X^{-1}}{N_{\al}N_{\ell}}\sum_{y}^{X}\left(\sum_{\al\alp}^{\phi N_{\al}}\lganav{\sum_{\ell}^{N_{\ell}} e^{i\vq\cdot\vecr_{\ell}^{\al}}}\lganav{\sum_{\ellp}^{N_{\ell}}e^{-i\vk\cdot\vecr_{\ellp}^{\alp}}}\right)_{\!\!y},\label{SDR2}
\end{align}
\end{subequations}
where $y$ indexes disorder realisations and the correlation functions are calculated within each realisation. One way to circumvent excessive summing involves supposing that we can obtain equivalent results by taking a single distribution of cross-links and summing correlation functions over different ways of labeling the system,
\begin{equation}\label{SumLabelingPermutations}
\frac{1}{X}\sum_{y}^{X}\left(\sum_{\al\alp}^{\phi N_{\al}} \anav{\dots}\right)_{\!\!y}
\to
\frac{1}{\binom{N_{\al}}{\phi N_{\al}}}\sum_{x}^{\binom{N_{\al}}{\phi N_{\al}}}\sum_{\al\alp\in x}^{\phi N_{\al}}\anav{\dots},
\end{equation}
where $x$ enumerates ways of choosing labels. The indices $\al$ and $\alp$ are then taken from the set specified by $x$. In terms of scattering measurements we should be able to sum over exactly the same functions this way, given a large enough system. Nevertheless, we are still obliged to sum up over the fraction of labeled monomers as many times as there are ways of choosing those labels and this is unnecessarily computationally exhaustive. Fortunately this method leads to a more efficient means of calculating these averages. The sum over different ways to choose $\phi N_{\al}$ labels can be rewritten as a weighted sum over same- and different-chain correlations between all possible $N_{\al}$ labels,
\begin{subequations}\label{SumAllCorrelations}
\begin{multline}
\scS_{0}(\vq)\delta_{\vq\vk}
=
\frac{\phi}{N_{\al}N_{\ell}}\left\{\sum_{\al}^{N_{\al}}\lganav{\sum_{\ell\ellp}^{N_{\ell}} e^{i(\vq\cdot\vecr_{\ell}^{\al}-\vk\cdot\vecr_{\ellp}^{\al})}}\right.
\\
\left.+\frac{\phi N_{\al}-1}{N_{\al}-1}\sum_{\al}^{N_{\al}}\psum{\alp\ne\al}{{N}_{\al}}\lganav{\sum_{\ell\ellp}^{N_{\ell}} e^{i(\vq\cdot\vecr_{\ell}^{\al}-\vk\cdot\vecr_{\ellp}^{\alp})}}\right\}
\end{multline}
\begin{multline}
\Gamma_{0}(\vq)\delta_{\vq\vk}
=
\frac{\phi}{N_{\al}N_{\ell}}\left\{\sum_{\al}^{N_{\al}}\lganav{\sum_{\ell}^{N_{\ell}} e^{i\vq\cdot\vecr_{\ell}^{\al}}}\lganav{\sum_{\ellp}^{N_{\ell}}e^{-i\vk\cdot\vecr_{\ellp}^{\al}}}\right.
\\
\left.+\frac{\phi N_{\al}-1}{N_{\al}-1}\sum_{\al}^{N_{\al}}\psum{\alp\ne\al}{{N}_{\al}}\lganav{\sum_{\ell}^{N_{\ell}} e^{i\vq\cdot\vecr_{\ell}^{\al}}}\lganav{\sum_{\ellp}^{N_{\ell}} e^{-i\vk\cdot\vecr_{\ellp}^{\alp}}}\right\}
\end{multline}
\end{subequations}
where the primed sum indicates that each $\ell$ and $\ell'$ do not appear on the same chain. This drastically simplifies computation, since one need only consider a single realisation of connective and topological disorder. The calculation of all possible pair correlations within this realisation yields the disorder-averaged correlations appearing in Eqs.~(\ref{DisorderAveragedCorrelations}).

\section{Replica Random Phase Approximation}
\label{Sec:2}

We now construct a formalism appropriate for calculating disorder-averaged observables of heterogeneous systems, using those obtained from simulating homogeneous systems. As in other systems characterised by quenched disorder, we face the task of averaging over the logarithm of a disorder-dependent partition function. To accomplish this, we employ the techniques and language of replicas \cite{ea75}, well-known from spin glass theory \cite{cc05}. This approach has been used extensively in microscopic formulations of analogous polymer systems and is particularly convenient in our case \cite{bv92a,bv92b}.

The problem amounts to one of taking the disorder average of a given observable, $\anav{A}$. This quantity is itself thermally averaged via an integral over fluctuating density variables, with Boltzmann weight determined by a Hamiltonian, $\scH_{0}[\rho_{\vq}]$, such that the disorder average takes the form,
\begin{equation}\label{WeightedAvgObservable}
\quav{\anav{A}}
=
\quav{Z_{0}^{-1}\int\scD\rho_{\vq}~A[\rho_{\vq}]\exp{(-\beta\scH_{0}[\rho_{\vq}])}},
\end{equation}
where,
\begin{equation}\label{PartitionFunction}
Z_{0}
=
\int\scD\rho_{\vq}~\exp{(-\beta\scH_{0}[\rho_{\vq}])}.
\end{equation}
The implicit dependence of the normalising partition function, $Z_{0}$, on the distribution of quenched cross-links poses an analytical challenge. By removing it from the disorder average, we would be wrongly assuming that the quenched variables (namely, the cross-links) fluctuate on the same timescale as the density variables. Instead we rewrite Eq.~(\ref{WeightedAvgObservable}) as,
\begin{equation}\label{WeightedAvgObservableIntroduceLimit}
\quav{\anav{A}}
=
\lim_{n \to 0}\quav{Z_{0}^{n-1}\int\scD\rho_{\vq}~A[\rho_{\vq}]\exp{(-\beta\scH_{0}[\rho_{\vq}])}},
\end{equation}
and make the following identification,
\begin{equation}\label{ReplicatedPartitionFunction}
Z_{0}^{n}
=
\int\left(\prod_{a=1}^{n}\scD\rho_{\vq}^{a}\right)~\exp{\!\left(-\beta\sum_{a=1}^{n}\scH_{0}[\rho_{\vq}^{a}]\right)}.
\end{equation}
This is the essence of the replica trick and may be interpreted as an $n$-fold \emph{replication} of the system. It should be stressed that every Hamiltonian appearing in Eq.~(\ref{ReplicatedPartitionFunction}) corresponds to the same realisation of disorder. We can now write,
\begin{equation}\label{WeightedAvgObservableIntroduceReplicas}
\quav{\anav{A}}
=
\lim_{n \to 0}\int\left(\prod_{a=1}^{n}\scD\rho_{\vq}^{a}\right)~A[\rho_{\vq}^{1}]\,\quav{\exp{\!\left(-\beta\sum_{a=1}^{n}\scH_{0}[\rho_{\vq}^{a}]\right)}},
\end{equation}
where the choice of replica label in the functional argument of $A$ is arbitrary. We further simplify matters by introducing an effective Hamiltonian given by,
\begin{equation}\label{EffectiveHamiltonian}
\scH^{\eff}_{0}[\rho_{\vq}^{a}]
\equiv
-\beta^{-1}\ln{\!\left(\quav{e^{-\beta\sum_{a}^{n}\scH_{0}[\rho_{\vq}^{a} ]}}\right)},
\end{equation}
which describes the energy of the disorder-averaged system. Now the average over observable $A$ can be recast in the form,
\begin{equation}\label{WeightedAvgObservableWithHEff}
\quav{\anav{A}}
=
\lim_{n \to 0}\int\left(\prod_{a=1}^{n}\scD\rho_{\vq}^{a}\right)~A[\rho_{\vq}^{1}]\,\exp{(-\beta\scH^{\eff}_{0}[\rho_{\vq}^{a}])}.
\end{equation}
Note that the Hamiltonian $\scH_{0}[\rho_{\vq}^{a}]$ for a given realisation of quenched disorder is replica diagonal, whilst the effective Hamiltonian $\scH^{\eff}_{0}[\rho_{\vq}^{a}]$ retains interactions between different replicas. 

We can use this recipe for weighted averages to describe the density fluctuations calculated in our simulations \emph{after} a disorder average. In this parametrisation, Eqs.~(\ref{DisorderAveragedCorrelations}) can be rewritten as,
\begin{subequations}\label{EffectiveDisorderAveragedCorrelations}
\begin{align}
\effanav{\rho_{\vq}^{a}}{0}
&=
0,\label{EDAC1}\\
\effanav{\rho_{\vq}^{a}\rho_{\vq}^{\star a}}{0}
&=
N_{\al}N_{\ell}\scS_{0}(\vq),\label{EDAC2}\\
\effanav{\rho_{\vq}^{a}\rho_{\vq}^{\star\apr}}{0}
&=
N_{\al}N_{\ell}\Gamma_{0}(\vq),\quad a\ne\apr.\label{EDAC3}
\end{align}
\end{subequations}
for $\vq\ne\mathbf{0}$. We need only choose a reasonable Ansatz for $\scH^{\eff}_{0}[\rho_{\vq}^{a}]$ in order to recover these correlation functions using Eq.~(\ref{WeightedAvgObservableWithHEff}). Invoking the RPA, we assume Gaussian fluctuations in density. Under a further assumption of symmetry under permutation of replicas, we can uncouple wave vector space by coupling replicas to write,
\begin{equation}\label{GaussianHamiltonian}
\scH^{\eff}_{0}[\rho_{\vq}^{a}]
=
\frac{1}{2}\sum_{\vq}\sum_{a\apr}^{n}\rho_{\vq}^{a}\vK_{a\apr}(\vq)\rho_{\vq}^{\star\apr},
\end{equation}
with $[\vK^{-1}(\vq)]_{a\apr}=\effanav{\rho_{\vq}^{a}\rho_{\vq}^{\star\apr}}{0}$ such that,
\begin{equation}\label{BareInteractionMatrix}
[\vK^{-1}(\vq)]_{a\apr}
=
N_{\al}N_{\ell}(\delta_{a\apr}\,\scS_{0}(\vq) 
+
(1-\delta_{a\apr})\,\Gamma_{0}(\vq)).
\end{equation}
The inverse corresponding to Eq.~(\ref{BareInteractionMatrix}) yields an expression for the kernel in the form,
\begin{multline}\label{InverseBareInteractionMatrix}
\vK_{a\apr}(\vq)
=
\frac{\delta_{a\apr}}{N_{\al}N_{\ell}}\frac{\scS_{0}(\vq)+(n-2)\Gamma_{0}(\vq)}{(\scS_{0}(\vq)-\Gamma_{0}(\vq))(\scS_{0}(\vq)+(n-1)\Gamma_{0}(\vq))}
\\
-\frac{(1-\delta_{a\apr})}{N_{\al}N_{\ell}}\frac{\Gamma_{0}(\vq)}{(\scS_{0}(\vq)-\Gamma_{0}(\vq))(\scS_{0}(\vq)+(n-1)\Gamma_{0}(\vq))}.
\end{multline}

\spanfig{BareCorrelationDataPlot}{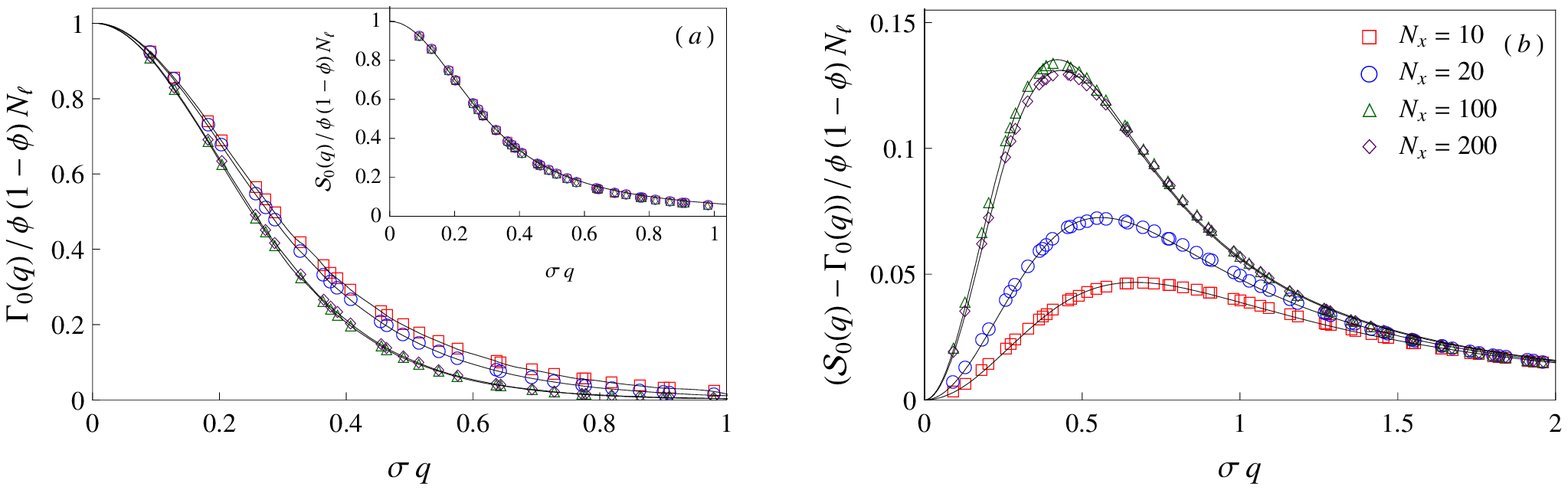}{55mm}{(a) Normalised structure function $\scS_{0}(\vq)$ (inset) and overlap function $\Gamma_{0}(\vq)$ data from molecular dynamics simulations of systems with different average cross-link separation $N_{x}$. Here, $\sigma$ is the length scale defined by the Lennard-Jones potential used in the simulations (see Appendix for details). Overlap data show strong dependence on $N_{x}$ when it is shorter than the entanglement length, $L_{e}=70$, whereas those for weakly cross-linked systems, as well as all structure function data, show no such dependence. Solid lines in main plot interpolate data. Solid line in inset plot corresponds to the Debye function. (b) Difference between normalised correlation functions $\scS_{0}(\vq)-\Gamma_{0}(\vq)$ peaks at characteristic wave vector $\qs$, which is dependent on cross-link separation, $N_{x}$.}

In order to investigate a heterogeneous analogue of this system, we now introduce an interaction potential, $V_{\vq}$ within the formalism. This inclusion is straightforward in the case for which the potential is a function of monomer {\it separation} only (and therefore diagonal in wave vector space). We perform the same formal treatment as for the homogeneous system, by first introducing the replica trick and then averaging with respect to disorder. The limit of vanishing replicas is trivial and physically reasonable at this stage of the calculation and in taking it we obtain the effective Hamiltonian,
\begin{align}\label{GaussianActionWithInteractions}
\scH_{V}^{\eff}[\rho_{\vq}^{a}]
&=
-\beta^{-1}\ln{\!\left(\quav{e^{-\beta\sum_{a}^{n}\scH_{V}[\rho_{\vq}^{a} ]}}\right)}\nonumber\\
&=
\scH_{0}^{\eff}[\rho]+\frac{1}{2\Omega}\sum_{\vq}\sum_{a=1}^{n}\rho_{\vq}^{a}V_{\vq}\rho_{\vq}^{\star a},
\end{align}
where the system is enclosed within volume $\Omega$, and,
\begin{equation}\label{PerturbedInteractionMatrix}
[ \vK(\vq)+\vV(\vq)]^{-1}_{a\apr}
=
N_{\al}N_{\ell}(\delta_{a\apr}\,\scS_{V}(\vq)+(1-\delta_{a\apr})\,\Gamma_{V}(\vq)),
\end{equation}
with $[\vV(\vq)]_{a\apr}\equiv\delta_{a\apr} V_{\vq}/\Omega$. Using the result from the homogeneous system in Eq.~(\ref{InverseBareInteractionMatrix}), we can rewrite Eq.~(\ref{PerturbedInteractionMatrix}) in the form,
\begin{multline}\label{NewPerturbedInteractionMatrix}
[ \vK(\vq)+\vV(\vq)]^{-1}_{a\apr}
=
N_{\al}N_{\ell}\delta_{a\apr}\times
\\
\frac{\scS_{0}(\vq)+\varrho V_{\vq}(\scS_{0}(\vq)-\Gamma_{0}(\vq))(\scS_{0}(\vq)+(n-1)\Gamma_{0}(\vq))}{1+\varrho V_{\vq}(\scS_{0}(\vq)-\Gamma_{0}(\vq)))(1+\varrho V_{\vq}(\scS_{0}(\vq)+(n-1)\Gamma_{0}(\vq))}
\\
+\frac{N_{\al}N_{\ell}(1-\delta_{a\apr})\Gamma_{0}(\vq)}{1+\varrho V_{\vq}(\scS_{0}(\vq)-\Gamma_{0}(\vq)))(1+\varrho V_{\vq}(\scS_{0}(\vq)+(n-1)\Gamma_{0}(\vq))},
\end{multline}
where $\varrho=N_{\al}N_{\ell}/\Omega$ defines the density of the system, and from this expression we can infer that,
\begin{subequations}\label{InteractingCorrelationFunctions}
\begin{multline}\label{InteractingStructureFunction}
\scS_{V}(\vq)
=
\frac{\scS_{0}(\vq)-\Gamma_{0}(\vq)}{1+\varrho V_{\vq}(\scS_{0}(\vq)-\Gamma_{0}(\vq))}
\\
+\frac{\Gamma_{0}(\vq)}{(1+\varrho V_{\vq}(\scS_{0}(\vq)-\Gamma_{0}(\vq)))^{2}},
\end{multline}
\begin{equation}\label{InteractingOverlapFunction}
\Gamma_{V}(\vq)
=
\frac{\Gamma_{0}(\vq)}{(1+\varrho V_{\vq}(\scS_{0}(\vq)-\Gamma_{0}(\vq)))^{2}},
\end{equation}
\end{subequations}
in the replica limit where $n\to0$.

The structure of the equations may be directly compared with results from microscopic descriptions of end-linked \cite{rbm95} and cross-linked \cite{wgz06} systems. The advantage with this formulation comes from the fact that we are free to explore material parameter space and deduce the behaviour of the system in different limits provided by the simulations. The derived equations clearly show the change effected in the behaviour of the system upon the introduction of inhomogeneities. The two heterogeneous correlation functions are formed by shifting their homogeneous counterparts by a self-energy term controlled by the interaction potential, $V_{\vq}$, which retains its functional form in Fourier space.

\section{Results}
\label{Sec:3}

The simulations provide data for several different parameter values and we can further tune our output using parameters arising from the post-processing. We begin by considering systems with label length $N_{\ell}=100$ and varying the number of cross-links, which in turn determines their average chemical separation, $N_{x}$. This variable in particular plays an important role in the behaviour of the overlap function, $\Gamma_{0}(\vq)$. In Fig.~\ref{BareCorrelationDataPlot}(a), we show data for the overlap function for different values of $N_{x}$. Data for $\scS_{0}(\vq)$ is inset here for comparison. We are restricted to a finite simulation box and thus cannot capture behaviour in the very long wavelength limit. However, we note the convergence of both correlation functions in this limit to finite values which we normalise to unity. It should be emphasised that different sets of  $\scS_{0}(\vq)$ data superimpose one another, implying that the data show no $N_{x}$-dependence. It is clear from the $\Gamma_{0}(\vq)$ data that the overlap is dependent on $N_{x}$, at least below a threshold of the order of the melt entanglement length of $L_{e}=70$ \cite{esgs04}.

We use the expression given in Eq.~(\ref{InteractingStructureFunction}) to post-process the data and obtain structure function data for the heterogeneous system. For local attractive interactions between labeled monomers, we set $V_{\vq}=-\chi$ and observe a divergence in $\scS_{V}(\vq)$ at finite wave vector where the difference $\scS_{0}(\vq)-\Gamma_{0}(\vq)$ forms a maximum. These maxima are illustrated in Fig.~\ref{BareCorrelationDataPlot}(b), for different values of $N_{x}$. In Fig.~\ref{StructureFunctionPlots} we show the post-processed data for different parameter values. Once again we note the convergence to a non-vanishing scattering intensity in line with early experiments \cite{bb88} and at odds with de Gennes original prediction \cite{pg-dg79a}. This discrepancy has been the subject of much investigation \cite{bvdb94,ss93,rbm95,lsb00,wzg05,wgz06}, but the long wavelength tendency in our simulations is unambiguously non-vanishing. 

In all plots in Fig.~\ref{StructureFunctionPlots}, the emergence of a characteristic length scale represented by a peak at finite $\vq$ may be interpreted as the establishment of pronounced spatial correlations between labeled chain densities. The position and size of peaks are strongly dependent on our position in parameter space. In each plot we vary a single parameter (interaction strength $\chi$, label fraction $\phi$ and average cross-link separation $N_{x}$ respectively) and examine the wave vector dependence of data leading towards divergence at characteristic values $\qs$ and $\chi_{S}=(\scS_{0}(\qs)-\Gamma_{0}(\qs))^{-1}$. In each case we also plot the value of $\qs$, which is obtained by fitting the appropriate $\scS_{0}(\vq)-\Gamma_{0}(\vq)$ curve and locating a maximum.
\putfig{StructureFunctionPlots}{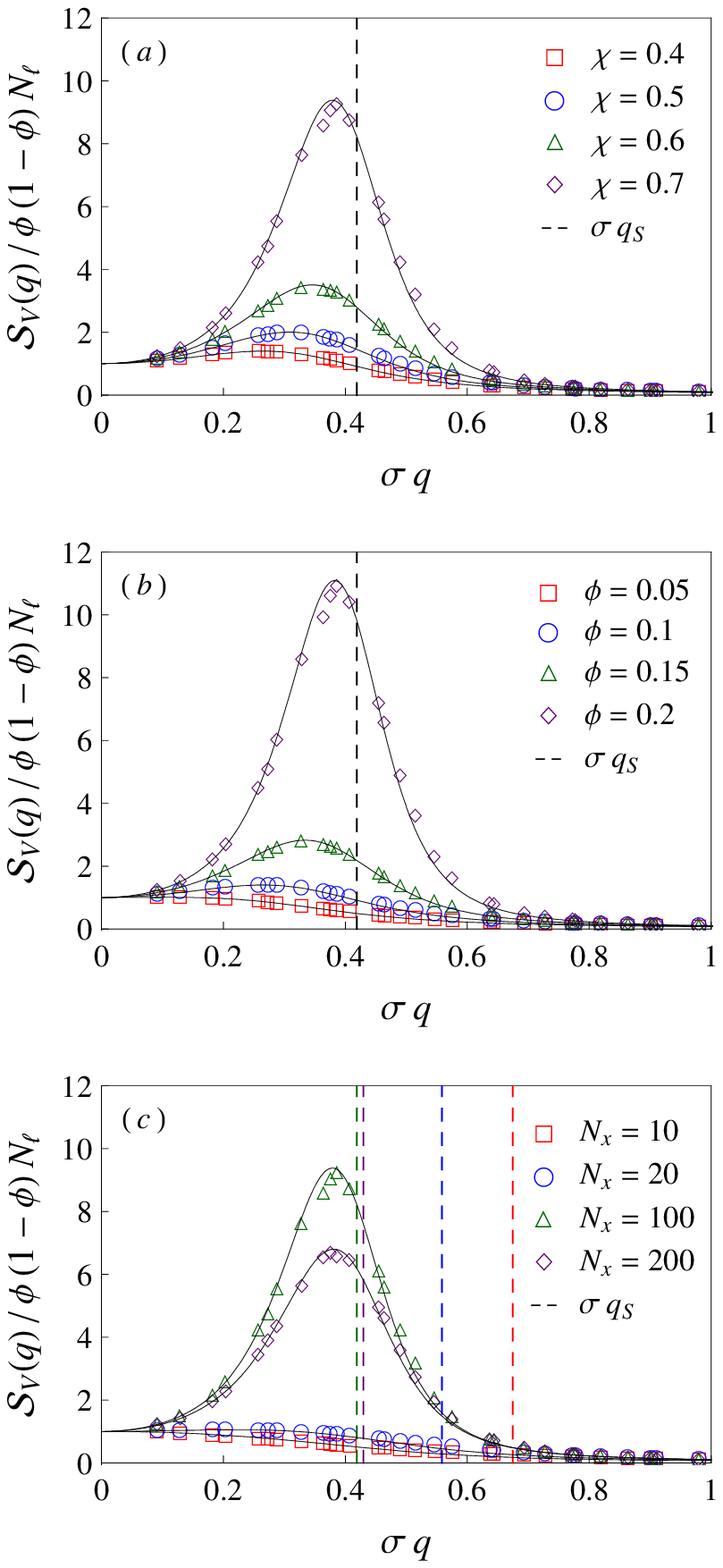}{180mm}{Post-processed structure functions for systems with (a) $\phi=0.1$, $N_{x}=100$, variable $\chi$; (b) $\chi=0.4$, $N_{x}=100$, variable $\phi$; and (c) $\phi=0.1$, $\chi=0.7$, variable $N_{x}$. Broken lines indicate the value of $\qs$ for respective parameters.}

In Fig.~\ref{StructureFunctionPlots}(a) we see the effect of increasing $\chi$ for a given fraction of labels and cross-links. There is a pronunciation of the peak leading to divergence and a corresponding shift in the wave vector value at which this peak occurs. A similar trend can be seen as the fraction of labels increases, as shown in Fig.~\ref{StructureFunctionPlots}(b). In this case, we see a characteristic length scale evident in systems with $20\%$ of polymers labeled for a given interaction strength. Under the same conditions, the structure function for the corresponding system with $5\%$ labeling shows no such tendency and instead the data resemble those of the homogeneous system. This disparity is more pronounced when comparing systems with different average cross-link separation, as seen in Fig.~\ref{StructureFunctionPlots}(c). We do not detect a divergence in strongly cross-linked systems, even as identical weakly cross-linked systems show a marked effect. Note that the inversion of data for $N_{x}=100$ and $N_{x}=200$ in Fig.~\ref{StructureFunctionPlots}(c) should be interpreted as an artifact of the simulation statistics.

\section{Discussion and conclusions}
\label{Sec:4}

We note that $\qs$ is independent of $\chi$ and $\phi$, the latter implying that the length scale describing a single label interacting with itself is indistinguishable from that describing several labels interacting. This in turn suggests that the divergence of the structure function is in fact a single-label effect. This observation is supported by other models but experimental and simulation results give the impression that more cooperative effects are prevalent in such systems \cite{lsb00}. However, we also note that the approach to separation is $\phi$-dependent, as shown in Fig.~\ref{StructureFunctionPlots}(b). This may be interpreted in the following way. The system is prepared in the initial homogeneous melt state and cross-links are formed. Simulated scattering measurements are indistinguishable from those of a polymeric liquid. Interactions are introduced between labels and the structure function is deformed slightly, but the physical properties remain largely unchanged. As interactions increase, some structure emerges at an intermediate length scale corresponding to a finite wave vector. Interactions are increased still further and labeled monomers become more attracted to one another, contracting the characteristic length scale and thus shifting the value of the corresponding wave vector. When the interaction strength reaches its spinodal value, the length scale of `domains' coincides with the intrinsic length scale of the system, related to the length scale of single-label collapse. The value of $\qs$ is independent of labeling fraction because the self-interacting length scale dominates the shape of the structure function.

Conversely, the value of $\qs$ is strongly dependent on $N_{x}$ when $N_{x}<L_{e}=70$ as illustrated by Figs.~\ref{StructureFunctionPlots}(c) and~\ref{QsNxDependencePlot}. In the latter, the dependence of $\qs$ on $N_{x}$ appears to obey an inverse square root law as seen in experiment \cite{bb88} and as predicted by classical rubber elasticity theory \cite{pg-dg79a}. This behaviour holds for different label lengths and breaks down for systems in which the average cross-link separation exceeds the entanglement length of the system, as indicated on the plot. We note in Fig.~\ref{DataCollapsePlot} that these data collapse when scaled with label length-dependent radii of gyration, $\lan R_{g}^{2}\ran(N_{\ell})$, and average cross-link confinement, $d_{X}$ \cite{sge05b,sge04}.
\putfig{QsNxDependencePlot}{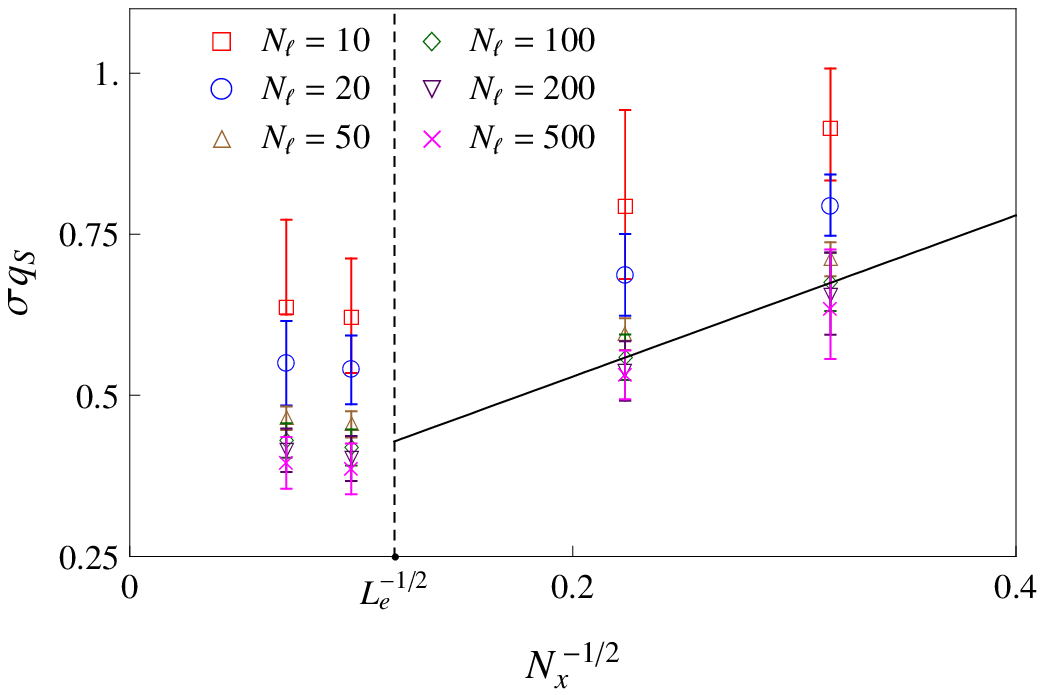}{57mm}{Dependence of spinodal wave vector $\qs$ on average cross-link separation $N_{x}$. Solid line fits $N_{\ell}=100$ data; broken line indicates length of entanglement, $L_{e}$. Again, $\sigma$ is the length scale defined by the Lennard-Jones potential used in simulations (see Appendix for details).}
\putfig{DataCollapsePlot}{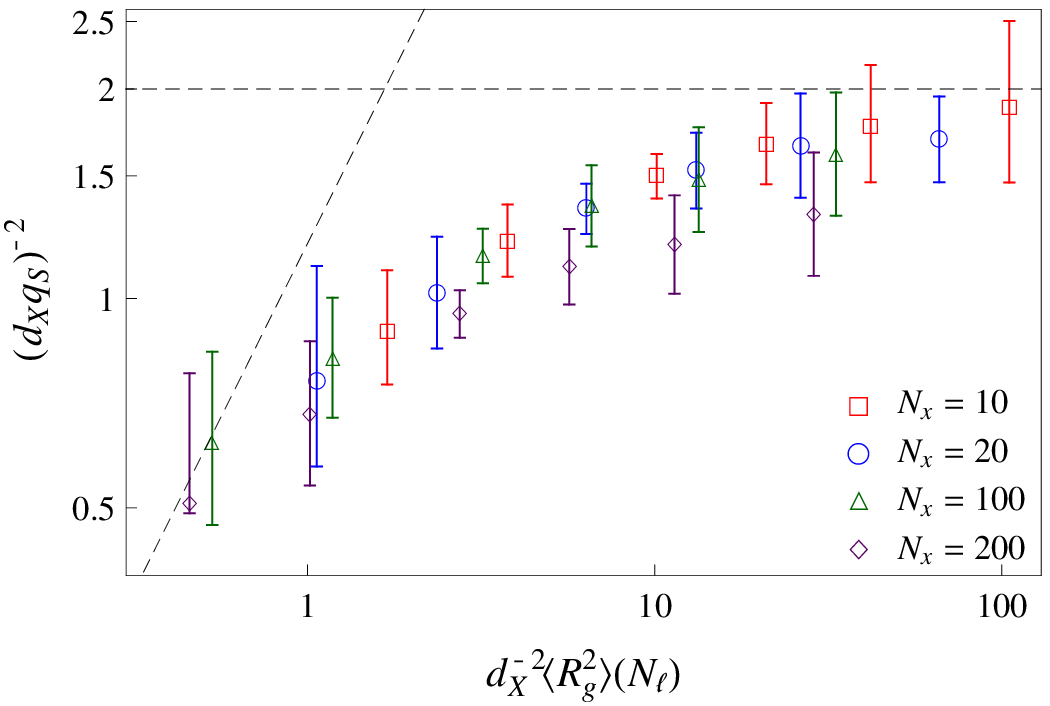}{55mm}{Data collapse when scaled with $N_{\ell}$-dependent radii of gyration, $\lan R_{g}^{2}\ran(N_{\ell})$, and average cross-link confinement, $d_{X}$.}

Finally, we find that we can use our post-processing formalism to determine exactly the character of the divergence for $\chi$ near $\chi_{S}$. The mean-field result predicting $(\chi_{S}-\chi)^{-1}$ behaviour was shown to describe experimental data well \cite{bb88}, but a  $(\chi_{S}-\chi)^{-2}$ divergence has been formally derived \cite{rbm95,wgz06}. We are able to locate the crossover from mean-field to higher order behaviour in the parameter space of our simulations. In analogy with previous studies \cite{pg-dg79a,bb88,rbm95}, we rewrite $\scS_{V}(\qs)$ in terms of the strength of the attractive interaction, $\chi$, as,
\begin{equation}\label{DivergentBehaviour}
\scS_{V}(\qs,\chi)
=
\frac{1}{\chi_{S}-\chi}+\frac{\chi_{S}^{2}\Gamma_{0}(\qs)}{(\chi_{S}-\chi)^{2}}.
\end{equation}
We define a parameter measuring `distance' from the divergent interaction strength, $\epsilon=|1-\chi/\chi_{S}|$, and note that the crossover occurs at some $\epsilon^{\star}$ given by,
\begin{equation}\label{DivergentCriterion}
\epsilon^{\star}=\left|1-\frac{\chi^{\star}}{\chi_{S}}\right|=\chi_{S}\Gamma_{0}(\qs).
\end{equation}
However, data from our simulations give $\epsilon^{\star}\in[0.66,0.74]$ for different values of $N_{x}$, indicating that higher order behaviour dominates our systems near the divergence. Experimental data \cite{bb88} showing a linear relationship between $\scS_{V}(\qs)^{-1}$ and an interaction parameter analogous to $\chi$ may not depict the truly divergent region of the systems' phase space, as illustrated by Fig.~\ref{DivergentBehaviourPlot}, and noted in Ref.~\cite{rbm95}. Here, we show that the linear relationship breaks down as $\chi$ approaches $\chi_{S}$. One may compare these curves with those in Fig. 7 of Ref.~\cite{bb88}. Our control over network characteristics via simulation proves particularly advantageous here; one can clearly see that the crossover to higher order behaviour occurs over a broader range of $\chi$ values in strongly cross-linked systems. This change is much harder to predict in more melt-like cases for which the average strand length exceeds that of entanglement.	
\putfig{DivergentBehaviourPlot}{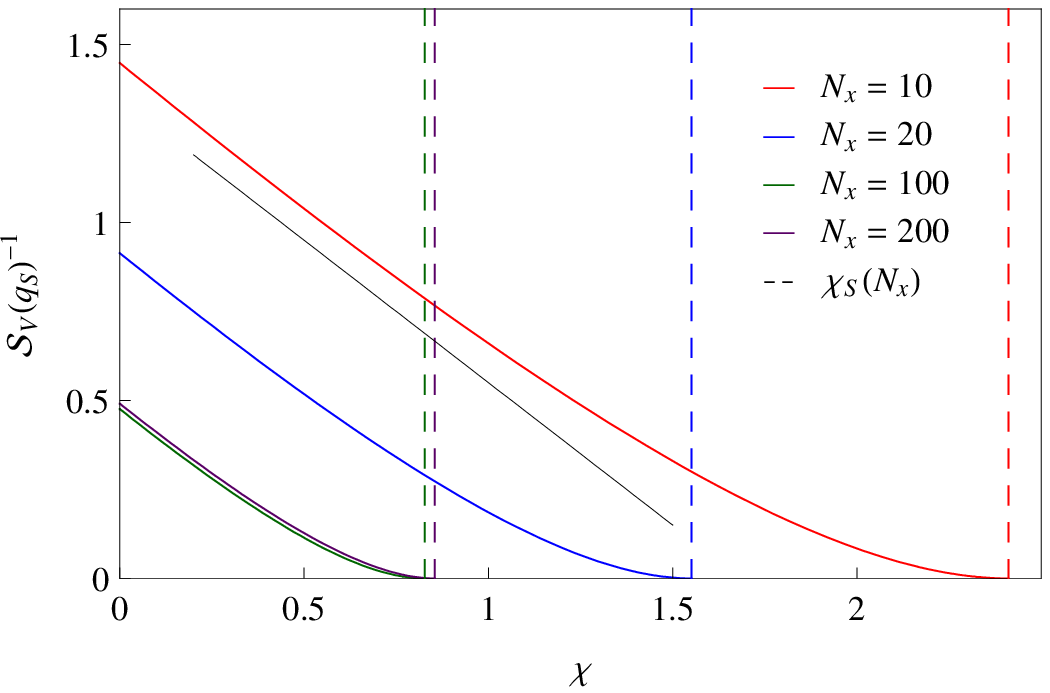}{55mm}{Crossover from $(\chi_{S}-\chi)^{-1}$ to $(\chi_{S}-\chi)^{-2}$ behaviour occurs as $\chi$ approaches $\chi_{S}$, indicated here by broken lines. A linear curve is plotted for comparison.}

To conclude, we have developed a means of examining microphase separation in polymer systems with quenched disorder without the need for exhaustive simulation. The formalism we use introduces heterogeneity in a controlled way and emerging results show a reproduction of experimental data. The observation of microphase separation in systems interacting via a localised potential calls for further analysis using these computational and analytical methods, particularly with respect to the accessible parameter space of the simulations. The preservation of the functional form of the interaction potential suggests the formalism may even be applicable to ferrogels \cite{jv04} and related systems involving non-local interactions.

\subsection*{Acknowledgements}
We would like to thank G.~Grest for molecular dynamics data. AVK thanks S.~Burdin for stimulating discussion and CS, J.~S.~Pedersen for insightful comments. CS gratefully acknowledges financial support from the Danish Natural Sciences Research Council through a Steno Research Assistant Professor fellowship. RE is supported by a Chair of Excellence grant from the Agence Nationale de Recherche (France).

\begin{appendix}
\section{Details of the simulations}
\label{app}

Polymer chains are modeled as sequences of beads connected by FENE springs,
while excluded volume interactions between beads are described by the shifted and truncated 12-6 Lennard-Jones potential,
\begin{equation}\label{FENE}
U_{\textrm{FENE}}(r)
=
-\frac{kR_{0}^{2}}{2}\ln\left( 1-\frac{r^{2}}{R_{0}^{2}}\right)\quad\textrm{for}\quad r<R_{0},
\end{equation}
\begin{equation}\label{LJPotential}
U_{\textrm{LJ}}(r)
=
4\epsilon\left(\left(\frac{\sigma}{r}\right)^{-12}-\left(\frac{\sigma}{r}\right)^{-6}+\frac{1}{4}\right)\quad\textrm{for}\quad r<2^{1/6}\sigma.
\end{equation}

The physical units for distance, energy and time are defined by parameters $\sigma,\epsilon$ and $\tau=\sigma\sqrt{m/\epsilon}$ respectively and all beads have mass $m=1$. The standard Kremer-Grest force field \cite{gk86,kg90} parameters have values $R_{0}=1.5\sigma$ and $k=30\epsilon\sigma^{-2}$. These parameter choices yield a high energy barrier for chain crossing which ensures conservation of the topological melt state. The Langevin dynamics is integrated using the LAMMPS code \cite{LAMMPS} with temperature $k_{B}T=1\epsilon$, monomeric friction $\Gamma=0.5 \tau^{-1}$ and time step $\Delta t=0.012\tau$. The resulting model polymers map onto those found in natural rubber \cite{sgke05}. 

Melts of 80 chains each with 3500 beads are generated with bead density $\varrho=0.85\sigma^{-3}$ and equilibrated as described in Ref. \cite{aegkp03}. The melts are instantaneously cross-linked by adding between $700$ and $14000$ FENE bonds between random pairs of beads separated by less than the reaction radius of $1.3\sigma$. This corresponds to nominal strand lengths $N_x$ between $10$ and $200$ monomers. The resulting randomly cross-linked networks are characterised by an exponential strand length distribution \cite{sge05b}. Table~\ref{NetworkCharacterisationTable} summarises our characterisation of the network. 
\begin{table}
\caption{Simulated network characterisation in terms of target strand length, $N_{x}$; effective strand length, $\lan N_{s}\ran$; effective functionality of cross-links, $\lan f\ran$; defect-free strand percentage, d.f.s.; and defect-free bead percentage, d.f.b., where dangling ends constitute a network defect.\label{NetworkCharacterisationTable}}
\begin{center}
\begin{tabular}{lllll}
\hline\noalign{\smallskip}
$N_{x}$ & $\lan N_{s}\ran$ & $\lan f\ran$ & d.f.s. & d.f.b.
\\
\noalign{\smallskip}\hline\noalign{\smallskip}
10 &
13 &
5.1 &
99\% &
99\%
\\
20 &
23 &
4.4 &
99\% &
99\%
\\
100 &
101 &
4.1 &
94\% &
93\%
\\
200 &
192 &
4.0 &
89\% &
87\%
\\
\noalign{\smallskip}\hline
\end{tabular}
\end{center}
\end{table}
\end{appendix}


\end{document}